%% file: main.tex
\documentclass[sigconf,10pt]{acmart}

\usepackage{times}
\usepackage{soul}
\usepackage{url}
\usepackage[utf8]{inputenc}
\usepackage{caption}
\usepackage{tikz}
\usepackage{amsmath,amsfonts,bm}
\usepackage{algorithm}
\usepackage{algorithmicx}
\usepackage{algpseudocode}

\usepackage{graphicx}
\usepackage{textcomp}
\usepackage{booktabs} 
\usepackage{siunitx}
\usepackage{xspace,url}

\usepackage{color}
\usepackage{colortbl}
\usepackage{multirow}
\usepackage{multicol}
\usepackage{makecell}
\usepackage{tabularx}
\usepackage{listings}
\usepackage{graphicx}
\usepackage{caption}
\usepackage{enumitem}
\usepackage{adjustbox}
\usepackage{subcaption}
\usepackage[colorinlistoftodos]{todonotes}

\acmSubmissionID{\#319}

\definecolor{airforceblue}{rgb}{0.36, 0.54, 0.66}
\newcommand{\name}{LLM-Explorer\xspace} 

\newcommand{\ie}{\textit{i}.\textit{e}.~}
\newcommand{\eg}{\textit{e}.\textit{g}.~}

\author{Shanhui Zhao$^{1,\dagger}$, Hao Wen$^{1,\dagger}$, Wenjie Du$^{1,2}$, Cheng Liang$^{1,3}$,\\ Yunxin Liu$^{1,5}$, Xiaozhou Ye$^{4}$, Ye Ouyang$^{4}$, Yuanchun Li$^{1,5,6,\ddagger}$}
\thanks{$\dagger$ Co-primary authors.}
\thanks{$\ddagger$ Corresponding author: Yuanchun Li (liyuanchun@air.tsinghua.edu.cn).}
\affiliation{$^1$~Institute for AI Industry Research (AIR), Tsinghua University \:$^2$~Hong Kong University of Science and Technology \:$^3$~Beijing University of Posts and Telecommunications \:$^4$~AsiaInfo Technologies (China), Inc \:$^5$~Shanghai Artificial Intelligence Laboratory
\:$^6$~Beijing Academy of Artificial Intelligence (BAAI)}

\renewcommand{\shortauthors}{Zhao et al.}

\begin{document}

\acmYear{2025}\copyrightyear{2025}
\acmConference[ACM MobiCom '25]{The 31st Annual International Conference on Mobile Computing and Networking}{November 3--7, 2025}{Hong Kong, China}
\acmBooktitle{The 31st Annual International Conference on Mobile Computing and Networking (ACM MobiCom '25), November 3--7, 2025, Hong Kong, China}
\acmDOI{10.1145/3680207.3723494}
\acmISBN{979-8-4007-1129-9/25/11}

\title[\name]{\name: Towards Efficient and Affordable LLM-based Exploration for Mobile Apps}




\begin{abstract}
Large language models (LLMs) have opened new opportunities for automated mobile app exploration, an important and challenging problem that used to suffer from the difficulty of generating meaningful UI interactions.
However, existing LLM-based exploration approaches rely heavily on LLMs to generate actions in almost every step, leading to a huge cost of token fees and computational resources.
We argue that such extensive usage of LLMs is neither necessary nor effective, since many actions during exploration do not require, or may even be biased by the abilities of LLMs.
Further, based on the insight that a precise and compact knowledge plays the central role for effective exploration, we introduce \name, a new exploration agent designed for efficiency and affordability. \name uses LLMs primarily for maintaining the knowledge instead of generating actions, and knowledge is used to guide action generation in a LLM-less manner.
Based on a comparison with 5 strong baselines on 20 typical apps, \name was able to achieve the fastest and highest coverage among all automated app explorers, with over 148x lower cost than the state-of-the-art LLM-based approach.
\end{abstract}



\maketitle

\input{tex/intro}
\input{tex/background}
\input{tex/approach}
\input{tex/experiment}
\input{tex/discussion}
\input{tex/conclusion}

\balance
\bibliographystyle{ACM-Reference-Format}
\bibliography{reference}

\end{document}

%% file: tex/intro.tex
\section{Introduction}

Automated mobile app exploration is a long-standing research problem, with lots of important applications including app testing \cite{droidbot, li2019humanoid, monkey, Caiipa, vfarms, fill_in_the_blank, liu2024vision, hu2024auitestagent}, malware detection \cite{mobisys_malware, mobicom_malware, cai2020assessing, LIReDroid}, and in-app data crawling \cite{tmc_app_crawling, collecting_data, macro_mining}.
The performance of mobile app exploration is usually measured by coverage, \ie the number of activities or lines of code reached in a limited period of time. Higher activity coverage correlates with better overall system performance. Several existing approaches, including Humanoid \cite{li2019humanoid}, GPTDroid \cite{gptdroid}, and DroidAgent \cite{droidagent}, have adopted this metric to evaluate their effectiveness.
Recently, emerging intelligent smartphone agents \cite{autodroid,memodroid,li2024personal} also rely on exploration to collect necessary knowledge for task automation.
Since the main interface of mobile apps is the graphical user interface (GUI or UI for short), the exploration of mobile apps is also usually grounded by GUI - The exploration agent navigates between different GUI states of an app by sending different GUI actions (touch, scroll, input text, etc.), just like how human users interact with the app.

The key question in mobile app exploration is how to generate the GUI actions that can efficiently discover new functionalities in the app. The major policies include random \cite{monkey, dynodroid, Sapienz} (randomly selecting which UI elements to interact with and how), model-based \cite{droidbot, test_study, droidmate, Poster_mobisys} (creating a model or representation of the UI and derive test cases to systematically cover different parts of the app functions), and learning-based \cite{li2019humanoid, deep_gui_ase, drl_test_ase} (leveraging machine learning or reinforcement learning techniques to generate test actions). Many explorers adopt a mixture of different policies.
Despite lots of existing attempts, there is still much room of improvement in app exploration. Mobile apps are dynamic, featuring a wide variety of user interfaces with diverse combinations of UI elements, and unique UI actions leading to different states. This complexity makes comprehensive exploration challenging, emphasizing the need for strategic testing methods \cite{missing_when_testing, vfarms, hotmobile_testing_service, llm_test_survey}. Therefore, we aim to efficiently cover as much app functionality as possible without needing to explore every single UI state, which might be infinite.
Achieving higher and faster coverage requires a deep understanding of the apps' functionalities and historical traces, which is difficult for most existing exploration agents.

Recently, pretrained foundation models, represented by large language models (LLMs), have demonstrated remarkable performance in language understanding, reasoning, and generation.
Which makes it possible for LLM-based agents to understand and execute tasks in a human-like fashion.
Since the app GUI can also be represented by natural language text and images, the LLM-based agents can potentially better understand the functionalities of apps and therefore improve the exploration performance.
Such an idea has already been studied in prior work \cite{llm_test_survey}. For example, GPTDroid \cite{gptdroid} introduces a method to generate testing actions by directly passing the current GUI information and historic trace to LLMs. 
DroidAgent \cite{droidagent} uses multiple autonomous agents for generating unexplored tasks, observing the interface, generating actions, and judging and reflecting on task completion. 
These approaches have demonstrated the ability of LLM to generate more meaningful GUI actions.

However, existing LLM-based exploration approaches have led to two important issues, limited efficiency and huge cost. 
The two issues originate from one cause - the excessive dependence on LLMs.
Existing approaches extensively query the LLMs to generate or plan steps, which is slow and costly (\eg generating a GUI action typically consumes around 500 tokens and 3 seconds, and exploring an app usually takes thousands of steps).
Actually, most steps during exploration do not demand much LLM-based reasoning and planning, which is analogous to human exploration of unknown places that is aimless at most times.
Moreover, asking LLMs to generate each action may bias the exploration process against unusual use cases, which are also important for exploration.

To address the above problems, we propose to combine the ability of LLMs and careful interaction knowledge modeling in mobile app exploration.
Our key insight is that \emph{the cornerstone of efficient exploration is the knowledge maintenance and utilization, instead of the action generation and planning abilities.} 
Specifically, similar to human or robots exploring a new environment, the key to efficient app exploration is reducing repetitive meaningless actions (\ie being creative), rather than planning future actions at each step (\ie being foresighted).
Since LLMs are trained by learning patterns from large datasets, they are usually considered to have only weak forms of creativity \cite{franceschelli2023creativity}.
On the contrary, an agent can be creative during exploration if it compares the candidate actions (and action combinations) against a high-quality knowledge. Maintaining the knowledge is mainly about information summarization, where LLMs can be more helpful.

Based on the insight, \name handles the automatic app exploration process with two main modules: LLM-assisted Knowledge Maintenance and Knowledge-guided Exploration. The \textbf{LLM-assisted Knowledge Maintenance} module is responsible for recording and categorizing reached UIs to prevent repetitive explorations or loops. This is crucial as variations in UI states or actions can make it challenging, even for LLMs. To address this, we introduce an app knowledge that contains abstract representations of UI states, elements and actions, as well as abstract interaction graphs to track transitions between UI states. The \textbf{Knowledge-guided Exploration} module then selects the next UI action to execute based on the maintained knowledge, with LLMs being occasionally invoked to tackle particularly complex UI actions (\eg text input). This approach aims to enhance exploration efficiency by reducing unnecessary LLM queries and focusing on strategic knowledge management and interaction planning.

We evaluate the effectiveness of our \name approach on 20 apps, in comparison with strong baselines including DroidAgent, GPTDroid, Humanoid, Droidbot, and Monkey. We also included the human exploration performance for reference.
The results have demonstrated that 
\name can achieve 4\%-35\% higher activity coverage than the baselines within a fixed time, matching human-level exploration efficiency on some apps. Compared with other LLM-based exploration approaches (DroidAgent and GPTDroid), \name can reduce the LLM cost by 9x-148x times. 

Our work makes the following technical contributions:
\begin{enumerate}
\item We study the LLM-based mobile app exploration with specific consideration on affordability, which can potentially benefit cost-sensitive individual mobile app developers and market-scale app analyzers.
\item We propose a new abstraction of an app's exploration knowledge, which is maintained by LLMs and can guide efficient future exploration. We believe this abstraction is also useful for a broader scope of app analysis.
\item Through a comprehensive evaluation, we demonstrate the effectiveness of our approach over strong baselines and the potential to advance the field of mobile app exploration.
\end{enumerate}

Our code is open-sourced at \url{https://github.com/MobileLLM/LLM-Explorer}.

%% file: tex/background.tex
\section{Background and Motivation}
\label{section:background}


\subsection{Mobile App Exploration}

App exploration is also called app traversal, crawling, or fuzzing based on different usage scenarios \cite{hotmobile_testing_service,tmc_app_crawling,droidbot}.
The goal of automated mobile app exploration is to traverse the functions of an app by automatically interacting with it. 
Since the main interface of mobile apps is GUI, the output of app exploration is usually a sequence of GUI actions, including touching, scrolling, typing text, etc.
Based on different purposes of app exploration, the output of exploration may also include the comprehensive report detailing the discovered bugs or issues, the performance metrics of the app, the data crawled from the app, or the screenshots or videos recording the exploration process.
More efficient and sufficient exploration usually leads to better performance in downstream applications, such as more bugs detected (for app testing), more precise malware classification (for malware detection), and more accurate task automation (for task automation).

The major performance metric of app exploration is the coverage. 
Unlike software testing approaches whose performance is usually measured with code coverage (\ie number of reached source code lines), mobile app explorers usually deal with apps without source code available.
The fundamental building block of an Android application is called activity, which represents a function component that provides a UI screen for users to interact with. For instance, an email app has different activities for different functions including the inbox, reading email, editing email, and settings. Therefore, measuring the effectiveness of an automated app explorer is often based on the activity coverage, \ie how many unique activities can be reached within a fixed time. 

Besides, implementing a mobile app explorer involves several key concepts about the GUI: 1) \textbf{UI state} refers to the current state of the mobile app visible through the user interface, which can interact with a user or an automatic explorer. 2) \textbf{UI element} is a visual component that users can interact with in a UI screen, including buttons, checkboxes, input boxes, etc. 3) \textbf{UI action} encompasses the interactions (such as clicks, text inputs, swipes) performed by users or automatic explorer to on UI elements to navigate or manipulate the app's functionality.
The job of an explorer is to send appropriate UI actions to navigate between different UI states.

\subsection{Difficulties of App Exploration}

Despite the numerous efforts in automatic app exploration, it faces fundamental challenges that hinder its further improvement. 
Existing app explorers can hardly match the efficiency and effectiveness of human users in traversing the app functions.
The main difficulties include: 

\textbf{1) Dynamic UI states.} Many apps can dynamically change content and layouts based on user data, interactions, or even preferences. For example, a contact list's UI may constantly evolve due to additions or modifications to the contacts, creating potentially infinite states. This variability can trap automatic explorers in endless loops if they fail to recognize these changes as the same state. 

\textbf{2) Large action space.} Some apps include a large number of interactive UI elements, a lot of which are similar or repetitive, (\eg time selectors, date pickers, checkbox list). This complexity makes comprehensive app exploration difficult, necessitating strategic testing plans to manage the extensive variety of options effectively. 

\textbf{3) Non-deterministic behaviors.} Mobile apps may exhibit unpredictable behaviors due to various factors, such as network latency, server-side processing, or interactions with other apps and services. These inconsistencies can make it challenging to navigate to specific UI states, hindering the explorer's ability to conduct thorough explorations.

\subsection{AI and LLM for Mobile App Exploration}

The aforementioned difficulties are mainly due to the lack of semantic understanding of GUI interactions, where AI techniques, especially LLMs can be helpful.
Machine learning models have the potential to enhance UI exploration efficiency by detecting dynamic UI changes and reducing repetitive actions. They excel in understanding UI component dependencies and relationships, enabling them to generate context-based test cases \cite{talking_with_ui, macro_mining}. 

However, most existing AI-based exploration methods, including deep learning \cite{li2019humanoid, deep_gui} and reinforcement learning approaches \cite{rl_testing, yasin2021droidbotx, rl_curiosity}, depend heavily on extensive training data and struggle with application generalization \cite{rico, li2019humanoid}. Despite various innovations, these methods often fall short of the human ability to quickly identify interactive elements, sometimes causing inefficiency or endless loops. LLMs, with their extensive pretraining on large-scale data and alignment with human preferences, present remarkable semantic understanding and zero-shot generalization ability for unseen apps and pages. This makes using LLMs a promising solution for mimicking human users' capabilities in app exploration~\cite{gptdroid,droidagent}.

Nevertheless, applying LLMs to mobile app exploration still presents several challenges. First, utilizing LLMs, whether deployed on the cloud or locally, incurs high inference costs. Accessing LLMs often involves considerable delays and expenses. Given that thoroughly exploring an app with high activity coverage typically needs numerous steps to explore an app (1,000+ steps at least), querying LLMs at each step leads to significant latency and costs. 
Second, LLMs could fall short in the creative thinking required for comprehensive app testing. They may tend to consistently select the main functions of the app based on its common sense while overlooking less apparent features \cite{franceschelli2023creativity}. This oversight can restrict the exploration of all app activities and hinder the discovery of critical issues.

%% file: tex/approach.tex
\section{Our Approach: \name}
\label{sec:approach}

\subsection{Overview}



\begin{figure*}
    \centering
    \includegraphics[width=0.85\textwidth]{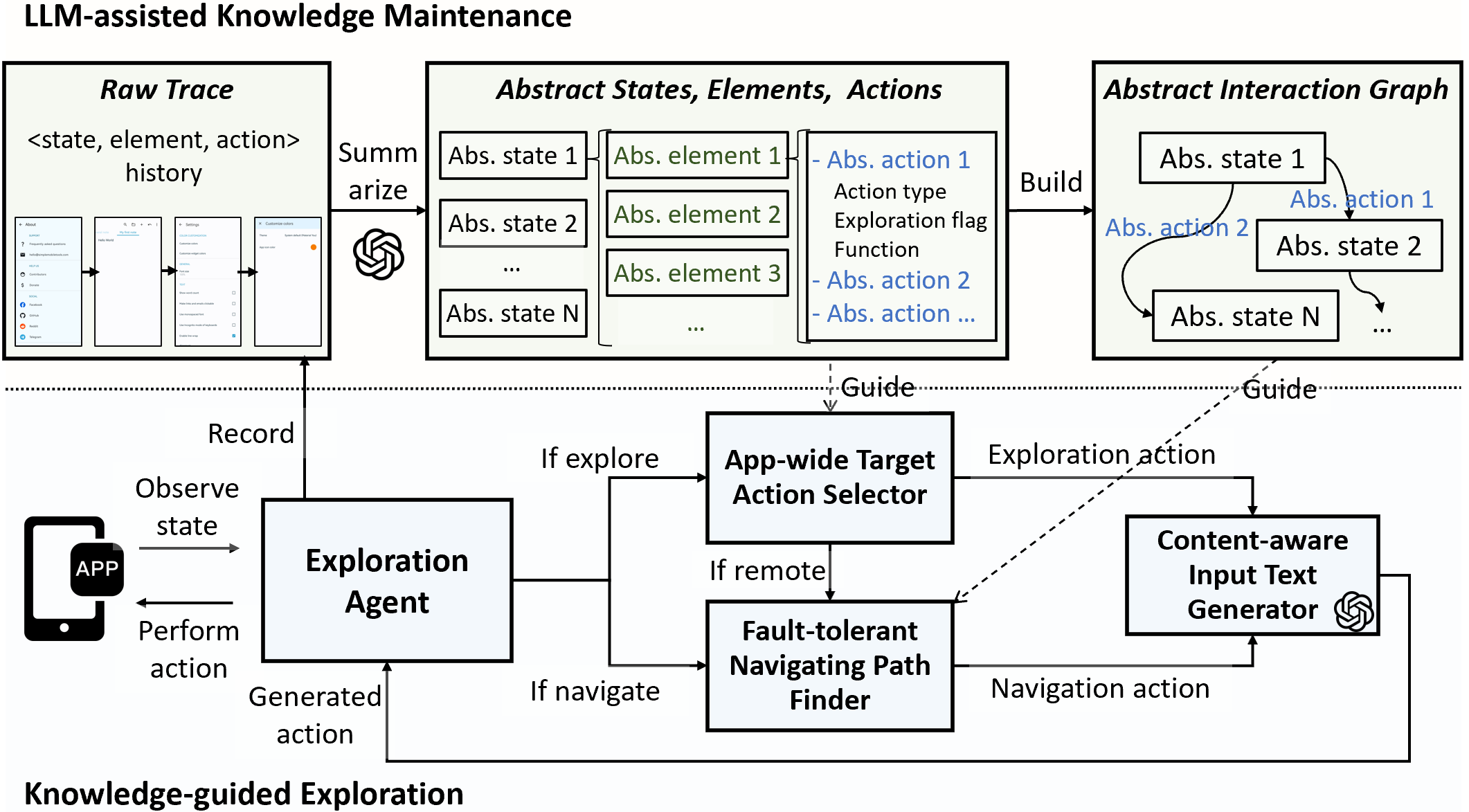}
    \vspace{-0.2cm}
    \caption{The workflow of \name.}
    \label{fig:overview}
    \vspace{-0.2cm}
\end{figure*}

We propose \name, an automated mobile application exploration system powered by LLMs to address the aforementioned challenges. The main idea of \name is to maintain a high-quality app knowledge to guide the exploration process. As shown in Figure \ref{fig:overview}, \name comprises two main modules: the LLM-assisted Knowledge Maintenance module and the Knowledge-guided Exploration module.

The LLM-assisted Knowledge Maintenance module utilizes LLMs to simplify the raw data of the exploration process, forming abstract app knowledge. It merges states and actions with similar functions into abstract states and actions, creating an Abstract Interaction Graph for smooth and efficient app navigation. The Knowledge-guided Exploration module utilizes the insights stored in the app knowledge, strategically choosing UI actions and devising test plans with context awareness. If the chosen action is not executable in the current UI state, \name navigates to the corresponding UI states using the Abstract Interaction Graph for guidance.

\subsection{LLM-assisted Knowledge Maintenance}

The input of the LLM-assisted Knowledge Maintenance module is a raw UI exploration trace, including a list of UI test steps featuring UI states, elements, and actions. The module maintains a high-quality app knowledge composed of abstract UI states, elements, actions, and an Abstract Interaction Graph (AIG). Prior approaches \cite{droidbot,li2019humanoid} also maintain a UI transition graph composed of raw states and actions, while we try to compress the knowledge components based on their semantic meanings. The idea is simple yet effective as it imitates how human users memorize concepts and reduce repetitive behaviors.



There are two problems to solve for maintaining the app knowledge: 
(i) Identifying UI states is crucial as seemingly identical UIs may represent different states, while dynamically changing UIs might belong to the same state. This distinction prevents excessive growth of app knowledge and aids in choosing efficient navigation paths. (ii) Reducing the UI action space is necessary because UIs can have many similar components, such as a calendar app's date elements for a month. Documenting every UI action increases redundancy and complicates the exploration module.





\subsubsection{Knowledge Organization}
\label{sec:mem_org}
\name builds up the exploration knowledge by summarizing the raw interaction trace.
Each step of the raw trace includes the UI state, the UI element in the state, and the UI action type (\texttt{touch}, \texttt{scroll}, etc.). 
To properly compress the knowledge, we introduce abstract UI states, elements, actions, and interaction graph in \name. 

An abstract UI state, symbolized as $s^{abs}_i$, represents a group of UI states ${s^{(1)}_i, s^{(2)}_i,...,s^{(j)}_i}$ encountered during exploration that serve the same set of functions within an app. The actual states of an abstract state may differ visually with different content (\eg two \emph{Contacts} screens with different contact names), but they are expected to share similar use cases.

Abstract UI actions group together similar user interactions, often invoking the same function or API call but with different parameters. 
This method identifies repetitive components within an application (such as dates or contact names) that usually result in the same abstract UI state upon interaction. 
By categorizing these interactions under a single abstract action, it simplifies the action space and avoids potential infinite loops. An abstract UI action is defined by four key components: \textit{Action Type}, \textit{Actual Element}, \textit{Exploration Flag}, and \textit{Function}:
1) \textit{Action Type} includes the four most common interactions: touch, long touch, scroll, and input. 
2) \textit{Actual Element} refers to specific UI elements involved in the action. These can either be a collection of elements within a single UI state that share similar functions or elements that are positioned identically across multiple actual states within an abstract state.
3) \textit{Exploration Flag} indicates the exploration status of an action with three possible values: unexplored (not yet executed), explored (has been executed by the exploration module), and ineffective (found to be non-functional or leading to no UI changes).
4) \textit{Function} denotes the functionality of the group of actual elements summarized by LLM.

The Abstract Interaction Graph (AIG) is a directed graph, whose nodes represent abstract UI states, and edges correspond to abstract UI actions. The source of the edge (abstract action) is the abstract UI state it is performed on, and the target is the resulting UI state after the action is performed. This model offers a more concise representation compared to the conventional UI Transition Graph (UTG) \cite{droidbot, macro_mining}. By omitting redundant actions and consolidating dynamic UI states, the AIG enables a more efficient navigation of UI states.


\subsubsection{Knowledge Update} 
\label{sec:mem_update}

\name updates the app knowledge after each exploration step. Each exploration step produces a tuple, representing the starting UI state ($s'$), the UI action taken ($a'$), and the ensuing UI state ($s$). 

Firstly, the abstract states are updated by matching the new UI state with existing states.
We attempted two methods to match the abstract UI states: the rule-based method and the LLM-based method.
The rule-based method capitalizes on the insight that UIs with similar functions, such as $s^{(m)}_i$ and $s^{(n)}_i$, often share comparable element structures. The differences lie in some dynamic UI element properties such as text content, selection status, scrolling positions, etc. Specifically, after excluding these dynamic properties, if two states $s^{(m)}_i$ and $s^{(n)}_i$ have the same set of elements, we combine them into a singular abstract UI state $s^{abs}_i$.
On the other hand, the LLM-based method queries LLMs to determine if two UI states are functionally equivalent. It is based on the former researches that prove LLMs to be effective for summarizing UI functions and answering UI-related questions \cite{talking_with_ui, autodroid}.
We found that using the rule-based method is more effective for merging the abstract states.
If the new UI state $s$ does not belong to either of the abstract UI state in the app knowledge, \name recognizes it as a new abstract UI state $s_i^{abs}$ and adds it to the app knowledge.

Secondly, \name updates the set of abstract UI actions. When encountering a new abstract UI state, \name aggregates possible UI actions that could be performed upon it into abstract UI actions $\{a_1^{abs}, a_2^{abs}...\}$, facilitated by querying LLMs. This is effective because LLMs excel at grouping repetitive UI components into generalized categories, thereby narrowing down the action space significantly. These newly incorporated abstract actions are labeled as $unexplored$. 
Meanwhile, for the action $a$ that has just been taken, its status is updated based on the UI state it leads to. The abstract state resulting from this action is compared to the preceding state's abstract state. If the resulting abstract state matches the previous one, it indicates that action $a$ didn't change the state, and the action is marked as $ineffective$. Otherwise, it is marked as $explored$.


Thirdly, \name updates the Abstract Interaction Graph $G$. If the abstract UI state $s^{abs}$ of the current state is newly created, \name creates a new node for the current $s^{abs}$ in $G$. Subsequently, \name checks whether there is a directed edge from the abstract state $s^{' abs}$ of the last state to $s^{abs}$ in $G$ with the attribute of the executed action. If such an edge does not exist, it is added to the graph.

The whole process is depicted in Algorithm~\ref{algo:knowledge-update}.

\begin{algorithm}
\footnotesize
\caption{Knowledge Update Process of \name.}
\label{algo:knowledge-update}
\begin{algorithmic}[1]
\Require Existing knowledge $K$ (including raw trace $K.T$, abstract states $K.S$, abstract actions $K.A$, abstract interaction graph $K.G$), last state $s'$, last action $a'$, and new state $s$
\Ensure Updated knowledge

\Function{UpdateKnowledge}{$K$, $s'$, $a'$, $s$}:
	\State Add $(s', a', s)$ to the raw trace $K.T$
	\State Try classify $s$ to existing abstract states $K.S$
	\If{$s$ matches existing abstract states $K.S$}
		\State $s^{abs} \gets$ matched abstract state of $s$ in $K.S$
	\Else
		\State $s^{abs} \gets$ create new abstract state from $s$
		\State Add $s^{abs}$ into $K.S$
	\EndIf
	\For{each new action $a$ in state $s$} \Comment LLM assisted
		\If{$a$ doesn't match existing actions $K.A$}
			\State $a^{abs} \gets$ create new abstract action from $a$
			\State $a^{abs}.exploration\_flag \gets$ \texttt{unexplored}
			\State Add $a^{abs}$ into $K.A$
		\EndIf
	\EndFor
	\State $s'^{\ abs} \gets$ matched abstract state of $s'$ in $K.S$
	\State $a'^{\ abs} \gets$ matched abstract action of $a'$ in $K.A$
	\State $a'^{\ abs}.exploration\_flag \gets$ \texttt{explored}
	\If{$s'^{\ abs} = s^{abs}$}
		\State $a'^{\ abs}.exploration\_flag.add(\texttt{ineffective})$
	\EndIf
	
	\State Update graph $K.G$ with transition $(s'^{\ abs}, a'^{\ abs}, s^{abs})$
	\State \Return Updated knowledge $K$
\EndFunction
\end{algorithmic}
\end{algorithm}

\subsection{Knowledge-guided Exploration Policy}


\subsubsection{Policy Overview}

Algorithm~\ref{algo:app-exploration} illustrates the exploration process. 
At the beginning, \name analyzes the current UI state and initializes the app's knowledge with it. It then enters a loop where it continuously executes actions, and updates the knowledge. This process continues until all abstract actions in the app's knowledge have been explored, marking the end of the exploration.

At the start of each explore step, the exploration agent first chooses an abstract action from the app knowledge to be executed based on the app-wide action selection method. 
Next, the agent verifies if the target element of the chosen action is present in the current UI state of the app. If it is, the action is executed immediately. If not, \name navigates to the UI state where the target UI element is located based on the fault-tolerant navigation path finder. 
After executing the UI action, \name updates the app knowledge with the new state transition, represented as $(s_i, a_i, s_{i+1})$.

\begin{algorithm}
\footnotesize
	\caption{Exploration Policy of \name.}
	\label{algo:app-exploration}
	\begin{algorithmic}[1]
\Require App under test, the device for testing
\Ensure Exploration knowledge and log

\Function{Explore-Main}{$app$}:
	\State $s_i \gets$ Observe GUI state of the $app$  \Comment Get initial state
	\State $K \gets UpdateKnowledge(\emptyset, null, null, s_i)$  \Comment Initialize knowledge
	\State $nav\_steps \gets \emptyset$  \Comment Initialize navigation steps
	\While{$K.unexplored\_abstract\_actions \neq \emptyset$}
		\If{$nav\_steps \neq \emptyset$} \Comment Try navigation
			\State $a_i \gets$ pop next action from $nav\_steps$ 
		\Else \Comment Try exploration
			\State $a_i \gets SelectExploreAction(K, s_i)$ 
			\If{$a_i \notin s_i.available\_actions$} \Comment Switch to navigation
				\State $nav\_steps \gets FindNavigatePath(K, s_i, a_i)$
				\State $a_i \gets$ pop next action from $nav\_steps$ 
			\EndIf
		\EndIf
		\If{$a_i$ is a text input action}
			\State $a_i \gets GenerateInputText(s_i, a_i)$ \Comment LLM assisted 
		\EndIf
		\State Perform action $a_i$ on the device
		\State Observe new state $s_{i+1}$
		\State $K \gets UpdateKnowledge(K, s_i, a_i, s_{i+1})$
		\If{$a_i$ is navigation and $s_{i+1}$ doesn't match $nav\_steps$}
			\State $nav\_steps \gets UpdateNavigatePath(K, nav\_steps)$
		\EndIf
	\EndWhile
\EndFunction
\end{algorithmic}
\end{algorithm}

\subsubsection{App-wide Action Selector}

\name introduces a novel strategy that focuses on individual UI elements across the entire app. This differs significantly from the existing methods that focus on UI state granularity \cite{droidbot, li2019humanoid, gptdroid}. 
This strategy stems from the insight that exploring an app by its individual UI elements could be more effective than examining through its various UI states. 
This is because an application can potentially generate an infinite combination of UI states by mixing and matching different elements.
However, the number of distinct UI elements, aside from some minor variations like content descriptions or colors, remains limited.
As a result, the exploring algorithm could reach completion once it has examined all the UI elements. 
This is in contrast to UI state granularity methods, which may find themselves in continuous loops when faced with dynamic UI states. 
Consequently, at each exploration step, \name selects and executes an abstract UI action from the app knowledge, rather than selecting a UI state to investigate.

Using its high-quality app knowledge that encompasses all UI actions, \name is able to choose from all the UI actions encountered rather than being restricted to the actions available in the current UI state. 
\name starts by going through the abstract actions in app knowledge, pulling out actions available in the current UI state that are labeled as $unexplored$. 
This creates a set of UI actions, among which it randomly picks one as the next action to explore. 
If no unexplored actions are found in the current state, \name widens its search to all UI states, gathering all the unexplored elements and action types to form an action pool, from which it randomly selects the next action for exploration. 
Compared to the LLM agents that select actions only from the current UI state \cite{gptdroid, droidagent}, this strategy saves LLM queries because it does not query LLM when navigating to a specific UI state.

\subsubsection{Fault-tolerant Navigating Path Finder}

The exploration agent of \name should possess the capability to navigate to a specific UI state, especially when the required UI action is not available in the current state. This can be achieved by utilizing the abstract UI interaction graph, where the exploration agent can select the shortest path from the current to the target UI state where the UI action is available. In this graph, edges represent UI actions, allowing for sequential navigation. Should the shortest path fails, the system will attempt alternative paths for a limited number of times.

However, due to unpredictable app behavior, the target abstract UI state might sometimes be unreachable. In these cases, \name will restart the app and attempt to navigate from the initial state to the target UI state again. If this approach also fails, the navigation attempt will be deemed as failed, and the corresponding action will be removed from the interaction graph so that future navigations can avoid generating infeasible paths.

\subsubsection{Content-aware Input Text Generator}
Generating human-like input text for text boxes is complex, as it requires understanding how to use the mobile apps. Thus, we rely on LLMs for this task. \name generates a structured prompt based on the name of the app under test and the current state information, and then queries LLM to obtain 
the input text. An example is shown in Figure~\ref{fig:prompt_input}.


\subsection{Detailed Usage of LLM}
\label{sec:detailed_usage_of_LLM}

In this section, we further clarify the usage of LLM in our approach and analyze the cost. 
\name involves LLMs in two parts, including knowledge organization and text input generation.



\begin{figure}
    \centering
    \includegraphics[width=0.47\textwidth]{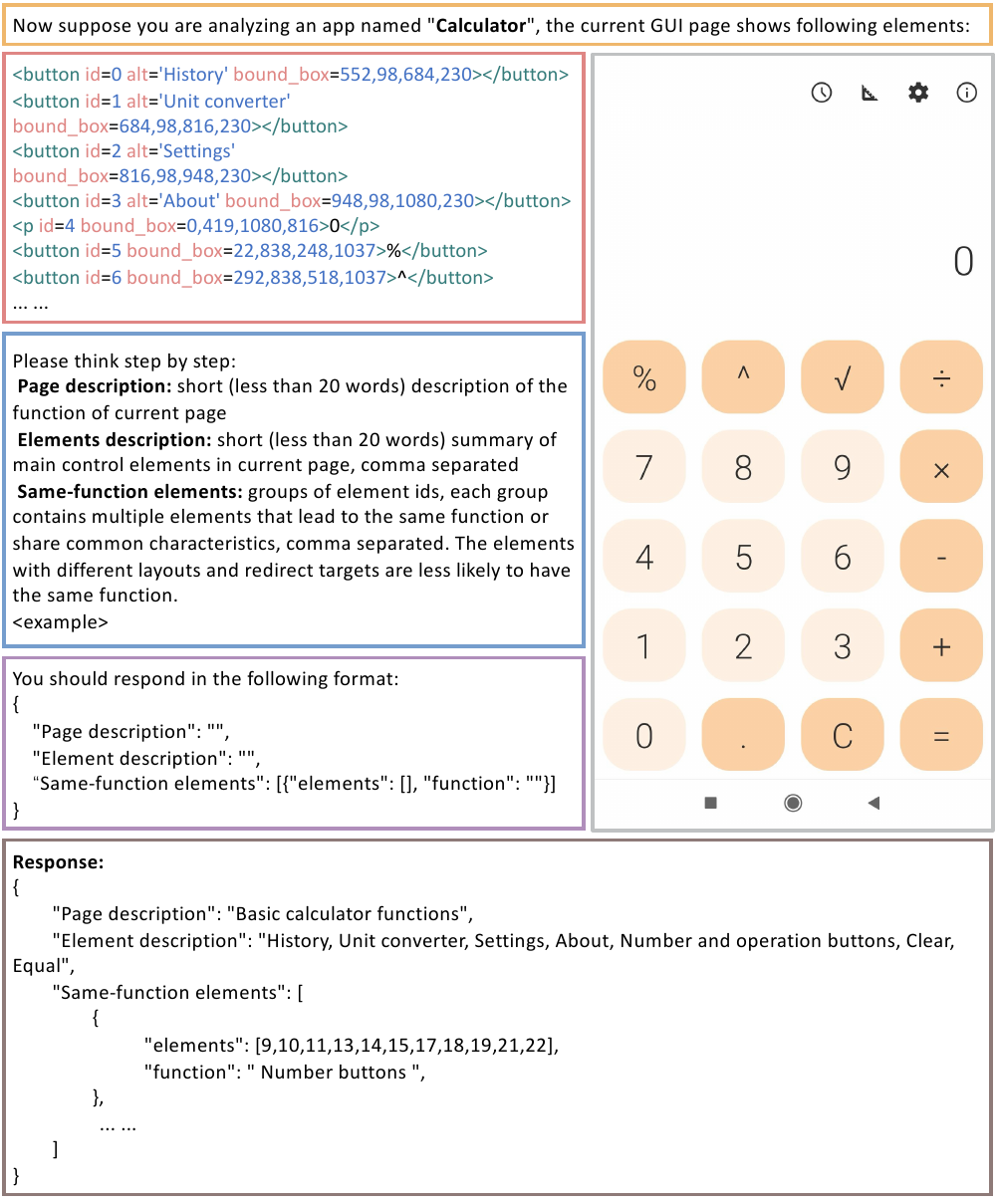}
    \vspace{-0.3cm}
    \caption{The prompt for knowledge organization. \textnormal{From top to bottom: general instructions, UI representation, chain-of-thought module, output format, and response, respectively.}
    }
    \label{fig:prompt_memory}
    \vspace{-0.2cm}
\end{figure}

\textbf{Using LLM for Knowledge Organization. }
The LLM is firstly utilized to categorize a group of similar UI actions, grouping them into an abstract UI action for app knowledge management, as detailed in Section \ref{sec:mem_update}. 
The structure of the prompt is illustrated in Figure \ref{fig:prompt_memory}, which is composed of four modules. 
It begins with an introduction including the testing app name and an explanation of the purpose of the prompt. 
Next, the UI is represented in the HTML syntax in line with previous studies \cite{talking_with_ui}. 
A chain-of-thought prompting module \cite{cot} is also included to encourage LLMs' logical reasoning. 
The output format is required to be a json dict for straightforward parsing. 
LLMs are expected to generate a merging instruction for UI elements, with the aim of consolidating elements with the same functions into a single UI element. 
This approach simplifies the action space by combining actions on these elements, which have similar functionality, into a single abstract UI action.



\begin{figure}
    \centering
    \includegraphics[width=0.47\textwidth]{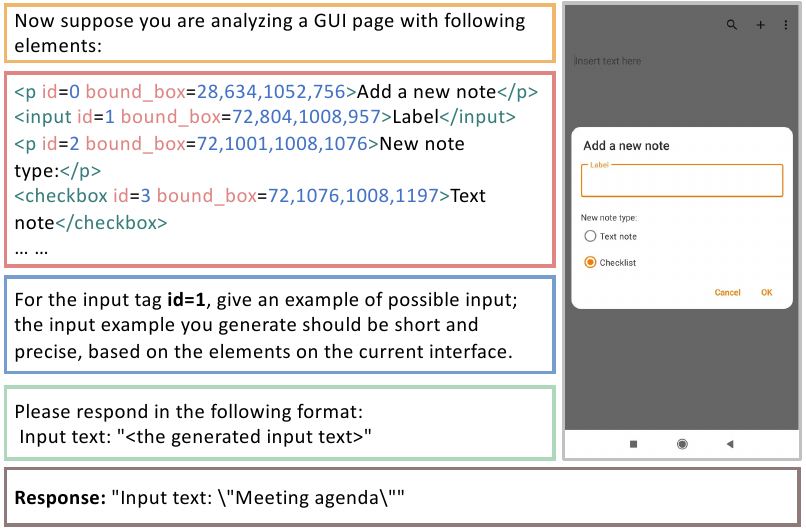}
    \vspace{-0.3cm}
    \caption{The prompt for content-aware input generation. \textnormal{From top to bottom, the prompt comprises: overall guidance, page representation, input request, response format, and response, respectively.}
    }
    \label{fig:prompt_input}
    \vspace{-0.2cm}
\end{figure}

\textbf{Using LLM for Content-Aware Input Generation.}
\name also employs the LLM to generate text inputs for UI elements, such as input boxes, based on the current UI context. Figure \ref{fig:prompt_input} illustrates the prompt structure. It provides the information about the current UI state and the specific input box to be filled and requests LLMs to generate the input text. We observe that LLMs could generate human-like input data, such as usernames, phone numbers, email addresses, etc. 

These two usages of LLM are uneasy for traditional AI techniques since it involves few-shot understanding of diverse UI content and free-form text generation.

According to the usages, the LLM queries of \name are determined by the number of unique states/actions (in knowledge organization) and the number of text input boxes (in input generation).
These numbers are bounded for most of the apps, although there might be a few exceptions with very dynamic GUI.
Meanwhile, these numbers are guaranteed to be smaller than the number of steps, since each unique state/action or input box maps to at least one step in the exploration.
Therefore, \name is expected to have much slower token consumption than existing approaches that use LLM for action generation.

%% file: tex/experiment.tex
\section{Evaluation}
\label{sec:experiment}

We implemented \name with Python and Java atop DroidBot \cite{droidbot} and GPT-3.5 (specific version: gpt-3.5-turbo-1106), and we conducted experiments on real app exploration tasks to evaluate its performance. In total, the experiments cost about \$1,000 for GPT services, including both GPT-3.5 and GPT-4 usage, covering \name and all baselines.  


\subsection{Experimental Setup}
\label{eval:setup}

\textbf{Benchmark Apps.}
We mainly evaluated \name on 20 apps collected from F-Droid and Google Play. The apps used in our study were selected based on the following rules: 1. Apps were chosen from the most common categories in app stores, with 1-3 apps selected per category. 2. Open-source versions of apps were prioritized where available. The sources and complexity of the apps is shown in Table \ref{tab:app_complexity}. We didn't include more apps for analysis due to the expensive cost of LLM services. However, we believe these apps are representative enough since they cover the most common app functionalities including contact management, messaging, photography, playing music, email management, and so on.

\begin{table*}[h]
\caption{Information about the benchmark apps}
    \centering
    \renewcommand{\arraystretch}{0.95}
    \begin{tabular}{c c c c || c c c c}
        \toprule
        \textbf{App Name} & \textbf{From} & \textbf{Activity Count} & \textbf{Complexity} & 
        \textbf{App Name} & \textbf{From} & \textbf{Activity Count} & \textbf{Complexity} \\
        \midrule
        Activity Diary & F-Droid & 11 & Medium & App Launcher & F-Droid & 8 & Low \\
        Calculator & F-Droid & 12 & Medium & Calendar & F-Droid & 18 & Medium \\
        Camera & F-Droid & 8 & Low & Clock & F-Droid & 14 & Medium \\
        Contacts & F-Droid & 12 & Medium & Draw & F-Droid & 8 & Low \\
        File Manager & F-Droid & 15 & Medium & Gallery & F-Droid & 23 & High \\
        Google Mail & Google Play & 64 & High & Keyboard & F-Droid & 10 & Low \\
        Markor & F-Droid & 11 & Medium & Music Player & F-Droid & 16 & Medium \\
        My Expenses & F-Droid & 53 & High & Notes & F-Droid & 10 & Low \\
        Open Tracks & F-Droid & 24 & High & SMS Messenger & F-Droid & 16 & Medium \\
        Voice Recorder & F-Droid & 12 & Medium & Wikipedia & F-Droid & 57 & High \\
        \bottomrule
    \end{tabular}
    \label{tab:app_complexity}
\end{table*}

\textbf{Hardware.}
We evaluate the end-to-end performance of \name on real Android devices and emulators with Android 10.
The exploration agents run on a desktop with 2 NVIDIA GeForce RTX 3090 GPUs and 48GB memory.

\textbf{Baselines.}
We chose DroidAgent \cite{droidagent}, GPTDroid \cite{gptdroid}, Humanoid \cite{li2019humanoid}, DroidBot \cite{droidbot}, and Monkey \cite{monkey} as our baselines.
DroidAgent and GPTDroid are LLM-based automated exploration methods. 
DroidAgent uses LLMs (GPT-4 and GPT-3.5) to plan tasks, observe the page, generate actions, determine task completion, etc. 
GPTDroid uses an LLM to choose the next action based on the current UI state.
Humanoid uses a deep neural network to generate human-like UI actions during exploration.
DroidBot is a generic app testing framework, in which we used the default depth-first traversal policy to explore the apps.
Monkey is a popular and official command-line tool for automated app fuzz testing.
Note that since the source code of GPTDroid is not fully usable at the time of our experiments, we use DroidAgent's reimplementation of GPTDroid, which uses a function-call-based action selector instead of the matching network in GPTDroid. We also use GPT-3.5 instead of GPT-3 in GPTDroid.
We also compared other variations of \name based on different LLMs (Vicuna-13B and GPT-4 (specific version: gpt-4-1106-preview) in Section~\ref{eval:ablation}.

\textbf{Reference Human Performance.}
Since certain app activities lack predefined navigation logic set by developers, the upper limit of activity coverage during exploration may not reach 100\%. Therefore, to better understand the performance of the automated explorers, we included human performance as a reference  for the upper bound of explorable activities. The human performance is collected with a user study approved by our Institutional Review Board (IRB) with the research question ``How many activities of apps can human users reach?''. We invited 6 experienced smartphone users to explore the 20 apps in Table \ref{tab:activity_coverage} using our lab device. Participants were selected based on their proficiency with computers and smartphones, with priority given to those with a background in computer science. They were all from our campus, had at least five years of smartphone experience, and possessed foundational knowledge in computer science.
The number of participants is determined by how many times each benchmark app can be explored. While we acknowledge the limited scale of the user study, we emphasize that six participants sufficed to explore each app twice. This provides adequate exploration coverage per app.
Participants were asked to explore all the pages and elements with different functions within the apps. The exploration time for each app was required to be no less than 10 minutes and the exploration of an app ended when the participant felt there was no more to explore.
During the process, our backend system automatically recorded the activities explored by the participants. In the end, each app was explored for more than twice, and we took the union of explored states as the final result.
To the best of our knowledge, this is also the first time in the community to compare against reference human performance in app exploration.

\textbf{Metrics.}
Similar to most existing work on mobile app exploration, we test our method and the baselines by comparing the progressive coverage, \ie the improvement of coverage over time and the achieved final coverage.
Since measuring the source code coverage is difficult for compiled APKs, we opt for monitoring the activity coverage (\ie the ratio of Activities reached by the agent among all Activities defined in the app), which is also a common practice.

\subsection{Exploration Effectiveness and Efficiency}
\label{eval:exploration_efficiency}


To evaluate the effectiveness and efficiency of \name, we test 20 popular apps with \name and the five baselines. To ensure fair comparisons, each method was given a fixed exploration time of 2 hours.

Table \ref{tab:activity_coverage} presents the final activity coverage achieved by each method. The results show that \name is able to achieve a higher activity coverage than all baseline methods. While there is still some gap to human performance, it can attain a similar or even higher level of coverage than human exploration on some simple apps.

\begin{table*}
\caption{The activity coverage achieved by \name and baselines on 20 typical apps.}
    \centering
    \renewcommand{\arraystretch}{0.95}
    {
        \begin{tabular}{c|cccccc|c}
		\toprule
            App      &  \name     &  DroidAgent      & GPTDroid   & Humanoid  &  DroidBot   & Monkey & Human\\
        \midrule
            Activity Diary & 90.91\% & \textbf{100.00\%} & 54.55\% & 90.91\% & 81.82\% & 45.45\% & 100.00\% \\
            App Launcher & \textbf{87.50\%} & \textbf{87.50\%} & 75.00\% & 75.00\% & 12.50\% & 37.50\% & 87.50\% \\
            Calculator & \textbf{83.33\%} & \textbf{83.33\%} & 50.00\% & 58.33\% & \textbf{83.33\%} & 25.00\% & 83.33\% \\
            Calendar & \textbf{61.11\%} & \textbf{61.11\%} & 22.22\% & 55.56\% & 33.33\% & 16.67\% & 61.11\% \\
            Camera & \textbf{87.50\%} & \textbf{87.50\%} & 75.00\% & 75.00\% & \textbf{87.50\%} & 50.00\% & 87.50\% \\
            Clock & \textbf{57.14\%} & 42.86\% & 28.57\% & \textbf{57.14\%} & 28.57\% & 35.71\% & 57.14\% \\
            Contacts & \textbf{75.00\%} & \textbf{75.00\%} & 41.67\% & 58.33\% & 50.00\% & 58.33\% & 83.33\% \\
            Draw & 62.50\% & \textbf{75.00\%} & 12.5\% & 62.50\% & \textbf{75.00\%} & 50.00\% & 75.00\% \\
            File Manager & \textbf{73.33\%} & 26.67\% & 26.67\% & 40.00\% & 53.33\% & 46.67\% & 53.33\% \\
            Gallery & \textbf{52.17\%} & 47.83\%  & 21.74\% & 44.00\% & 21.74\% & 13.04\% & 59.09\% \\
            Google Mail & \textbf{25.00\%} & 21.88\% & 4.69\% & 21.88\% & 7.81\% & 3.13\% & 28.13\% \\
            Keyboard & \textbf{70.00\%} & \textbf{70.00\%} & 10.00\% & \textbf{70.00\%} & \textbf{70.00\%} & \textbf{70.00\%} & 70.00\% \\
            Markor & \textbf{54.55\%} & 36.36\% & 36.36\% & 18.18\% & 36.36\% & 18.18\% & 54.55\% \\
            Music Player & 62.50\% & 62.50\% & 12.50\% & 56.25\% & \textbf{68.75\%} & 18.75\% & 75.00\% \\
            My Expenses & \textbf{26.42\%} & \textbf{26.42\%} & 7.55\% & 15.09\% & 13.21\% & 18.87\% & 52.38\% \\
            Notes & \textbf{80.00\%} & 70.00\% & 10.00\% & 70.00\% & 70.00\% & 10.00\% & 80.00\% \\
            Open Tracks & \textbf{66.67\%} & \textbf{66.67\%} & 29.17\% & 33.33\% & 37.50\% & 50.00\% & 50.00\% \\
            SMS Messenger & 81.25\% & \textbf{87.5\%} & 18.75\% & 43.75\% & 50.00\% & 25.00\% & 75.00\% \\
            Voice Recorder & \textbf{66.67\%} & 50.00\% & 25.00\% & 58.33\% & \textbf{66.67\%} & 25.00\% & 66.67\% \\
            Wikipedia & 28.07\% & 24.56\% & 10.53\% & \textbf{42.11\%} & 26.32\% & 7.02\% & 35.71\% \\
            \midrule
            Average & \textbf{64.58\%} & 60.13\% & 28.62\% & 52.28\% & 48.68\% & 31.22\% & 66.74\% \\
		\bottomrule
	\end{tabular}
    }
 \label{tab:activity_coverage}
\end{table*}

\begin{figure}
    \centering
    \includegraphics[width=0.47\textwidth]{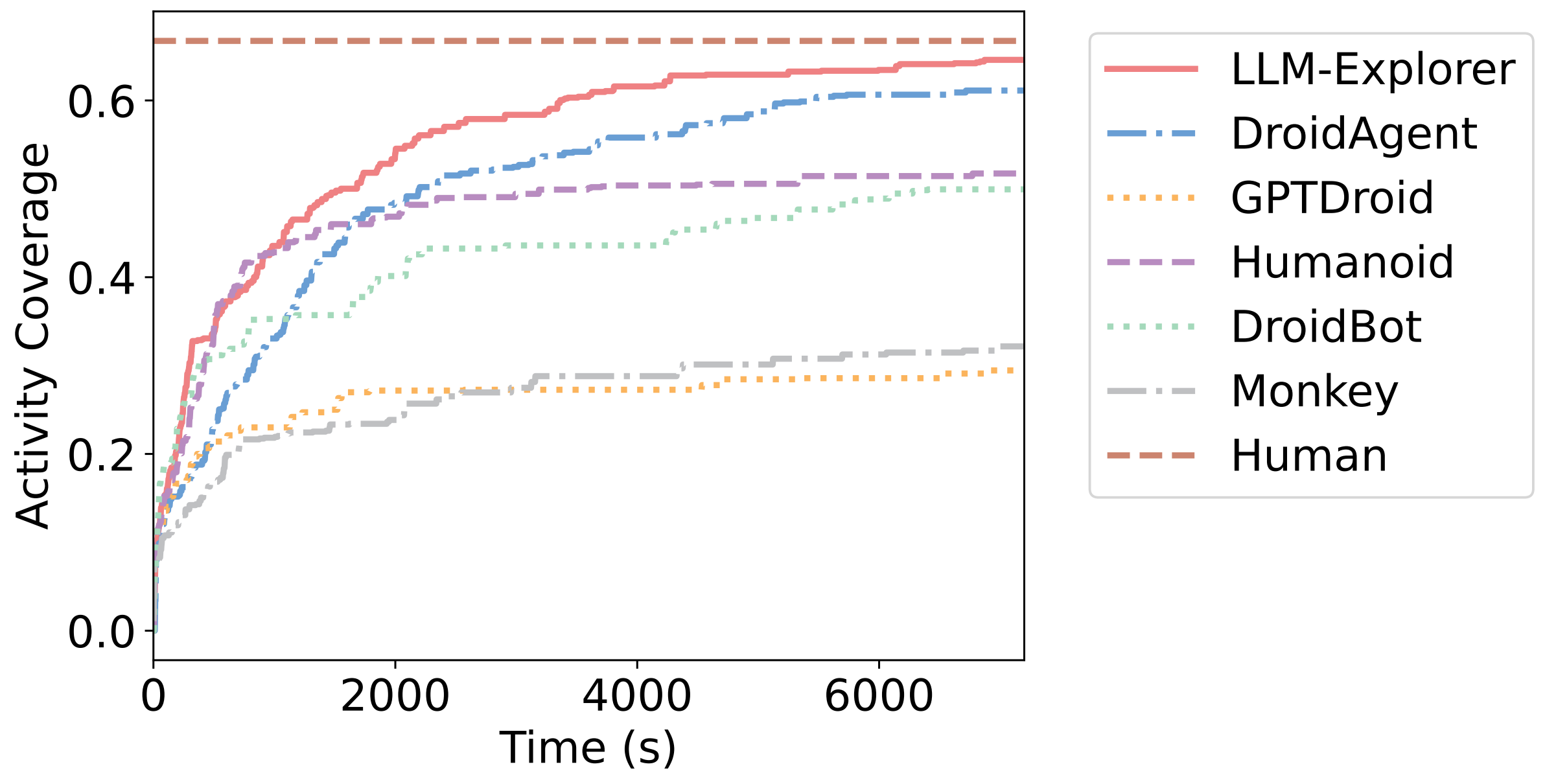}
    \caption{Progressive activity coverage of \name and baselines over time. \textnormal{The brown dotted line is the reference human performance.}}
    \label{fig:activity_coverage_by_time}
\end{figure}

\begin{figure}
    \centering
    \includegraphics[width=0.47\textwidth]{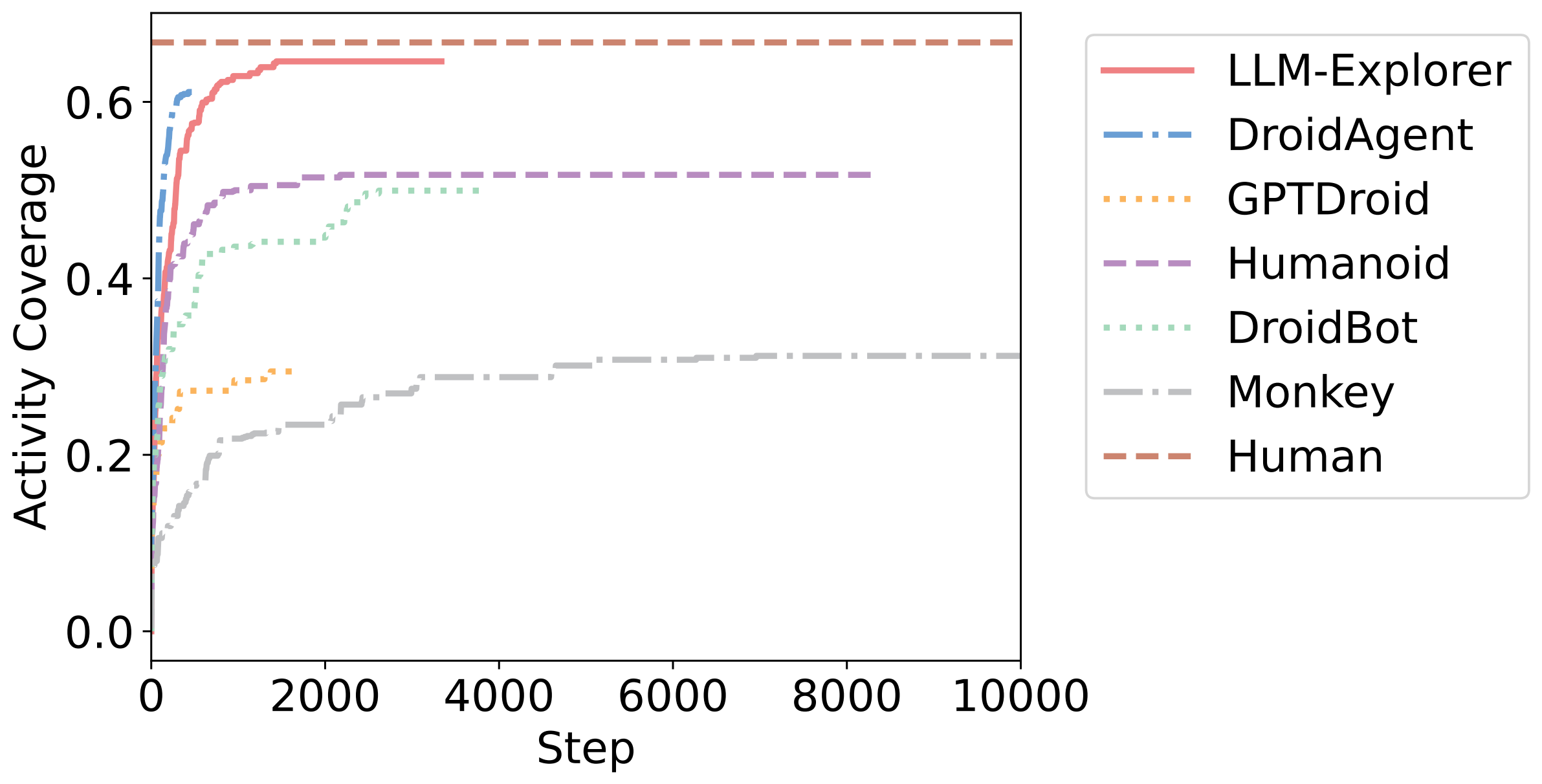}
    \caption{Progressive activity coverage of \name and baselines over steps within 2 hours. \textnormal{The brown dotted line is the reference human performance. The maximum step differs for each method due to the different per-step time. Note that Monkey produced over 20,000 steps in 2 hours of exploration, but the activity coverage nearly converged after 10,000 steps, so this figure only shows up to 10,000 steps.}}
    \label{fig:activity_coverage_by_step}
\end{figure}

Figure \ref{fig:activity_coverage_by_time} and Figure \ref{fig:activity_coverage_by_step} illustrate the progressive activity coverage of \name and baselines over time and over the number of steps, respectively.


As shown in Figure \ref{fig:activity_coverage_by_time}, at the beginning of exploration, \name, Humanoid, and DroidBot achieved similar coverage growth rates, which were significantly higher than DroidAgent, GPTDroid, and Monkey. This is because they could send actions at a faster speed and the actions could easily reach new activities in the early stage. Gradually, the growth rates slowed down as discovering new activities became harder, while \name and DroidAgent began to have higher growth rates than the others with the help of LLMs. Eventually, \name was able to achieve the highest activity coverage.

In Figure \ref{fig:activity_coverage_by_step}, the coverage of \name gradually converged at about 1500 steps, faster than most of the baselines. During exploration, although DroidAgent's per-step actions were more effective in contributing to activity coverage improvement, DroidAgent's each step was more time-consuming due to the fact that it requested LLM extensively for reasoning and planning, with the lowest number of actions generated in the 2-hour exploration. We also observed that the activity coverage growth rate of DroidAgent decreased with the number of steps at the end of exploration, which we thought was because LLM-based action selection became less effective for discovering more detailed and unusual activities. In Table \ref{tab:single_step_time}, we show the average single-step time of \name and baselines.


\begin{table}
 \caption{Average per-step time of \name and baselines. \textnormal{LE: \name, DA: DroidAgent, GD: GPTDroid, HU: Humanoid, DB: DroidBot, MO: Monkey.}}
    \centering
    \resizebox{.4\textwidth}{!}
    {
        \begin{tabular}{c|cccccc}
		\toprule
            Method      &  LE     &  DA      &  GD      & HU  &  DB   & MO \\
		\midrule
            Time (seconds)        &   5.19  &   19.10   &   5.59   &   3.19    &  2.94  &  0.79   \\
		\bottomrule
	\end{tabular}
    }
 \label{tab:single_step_time}
\end{table}

The advantages of \name over the baselines could be attributed to the following reasons: (i) \name merges states with identical functionality into a single abstract state, thereby eliminating redundant exploration of states with the same functionality. This strategy allows the exploration process to focus on discovering more unknown features of the app under test. 
Our further analysis found that each app has 128 abstract states on average. \name reduced the numbers of redundant state and action visits by 4.08 and 14.71 per abstract state, which effectively reduces the exploration space.
(ii) \name performs exploration based on abstract actions, which allows a more comprehensive coverage of all functionalities within the current app, including those that are not primary features, such as viewing the list of contributors on the ``About'' page of \emph{Notes} app. (iii) \name's LLM-assisted knowledge maintenance strategy of merging functionally identical elements avoids \name repeatedly exploring duplicate elements with the same functionality. For example, within the ``file import'' page of the \textit{File Manager} app, \name can combine the file name buttons on the page into the same abstract element. This makes it possible to explore the page by importing a file only once, thereby accomplishing the exploration of the file import functionality.

We further analyzed why \name failed in some cases: (i) \name merged different states into an abstract state based on rules, but if there was a unique element in the state, \name could not efficiently merge the state with the previous state. We also tried using LLM to merge states, but the result was not satisfactory, so we didn't adopt this solution in the end. (ii) In LLM-assisted knowledge maintenance, LLM may incorrectly treat elements with different functions as having the same function, resulting in the merging of elements that could reach different activities into one abstract element, causing \name to miss some activities when exploring. Examples are given in section \ref{sec:case_studies}.

\subsection{LLM Cost of Exploration}
\label{eval:llm_cost}


\name uses the LLM in two scenarios including knowledge maintenance and input text generation. GPTDroid and DroidAgent use LLMs mainly for action generation, and also for planning, observing, etc.
We compare the number of tokens and API fees consumed by different methods during the exploration process. Table~\ref{tab:average_cost} shows the average cost of using LLM to explore an app. As compared to DroidAgent (which achieved comparative coverage as ours), \name significantly reduced the average cost of exploring an app from \$16.31 to \$0.11, which was over 148 times less.

\begin{table}
 \caption{The average number of tokens and average cost consumed by \name, DroidAgent and GPTDroid when exploring an app. \textnormal{GPT-3.5, GPT-3.5-16K, and GPT-4 represent the tokens consumed when exploring an app using the corresponding model, respectively. Cost denotes the total cost of LLM when exploring an app.}}
    \vspace{-0.3cm}
    \centering
    \resizebox{.48\textwidth}{!}
    {
        \begin{tabular}{c|cccc}
		\toprule
            Method      &  GPT-3.5     &   GPT-3.5-16K     & GPT-4  &  Cost(\$)    \\
		\midrule
            \name        &   96,884.2  &   0   &   0    &  0.11   \\
            DroidAgent        &   1,384,079.0  &   1,141,095.29   &   335,913.29    &  16.31   \\
            GPTDroid        &   993,939.35  &   0   &   0    &  1.07   \\
            
		\bottomrule
	\end{tabular}
    }
    \vspace{-0.2cm}
 \label{tab:average_cost}
\end{table}




We also compared the number of query and the number of query tokens during exploration for different methods, as shown in Table \ref{tab:query_num_token}. \name reduces the number of query by 18.48 and 9.40 times compared to DroidAgent and GPTDroid respectively. At the same time, \name used fewer input tokens while obtaining approximately twice the number of output tokens compared to other methods, and also had the lowest average token count per query.

\begin{table}
 \caption{Number of queries and average number of query tokens by \name, DroidAgent and GPTDroid. \textnormal{}}
    \vspace{-0.3cm}
    \centering
    \resizebox{.48\textwidth}{!}
    {
        \begin{tabular}{c|ccc}
		\toprule
            Method      &  \name     &   DroidAgent     & GPTDroid    \\
		\midrule
            Number of queries        &   157.12  &   2903.15   &   1476.51   \\
            Input tokens per query        &   507.15  &   933.71   &   623.00   \\
            Output tokens per query        &   109.47  &   51.80   &   50.17   \\
            
		\bottomrule
	\end{tabular}
    }
 \label{tab:query_num_token}
\end{table}

\begin{figure}
    \centering
    \includegraphics[width=0.4\textwidth]{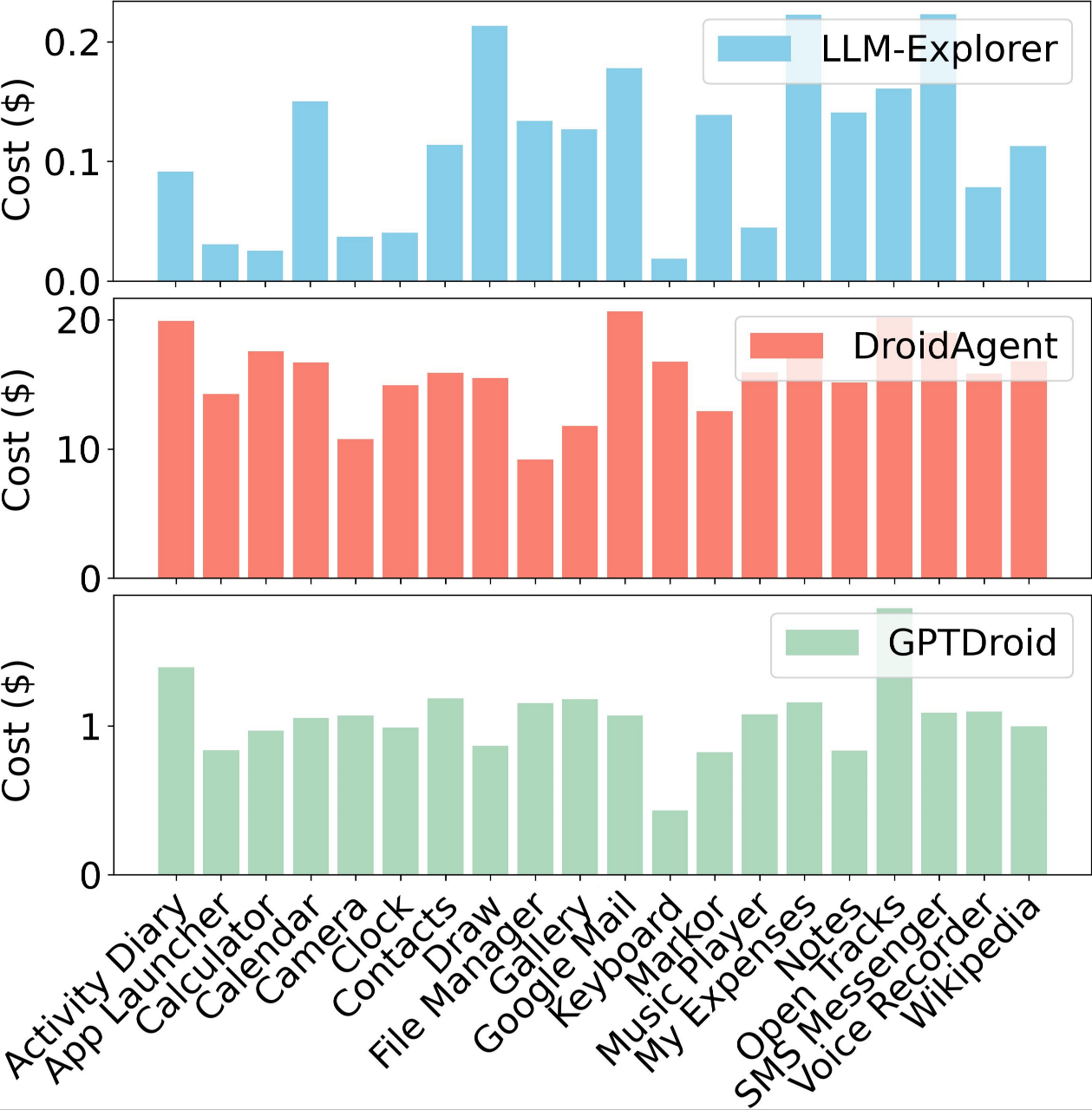}
    \caption{The cost of exploration in all apps under test by \name, DroidAgent, GPTDroid, respectively.}
    \label{fig:cost_apps}
    \vspace{-0.3cm}
\end{figure}

\begin{figure}
    \centering
    \includegraphics[width=0.33\textwidth]{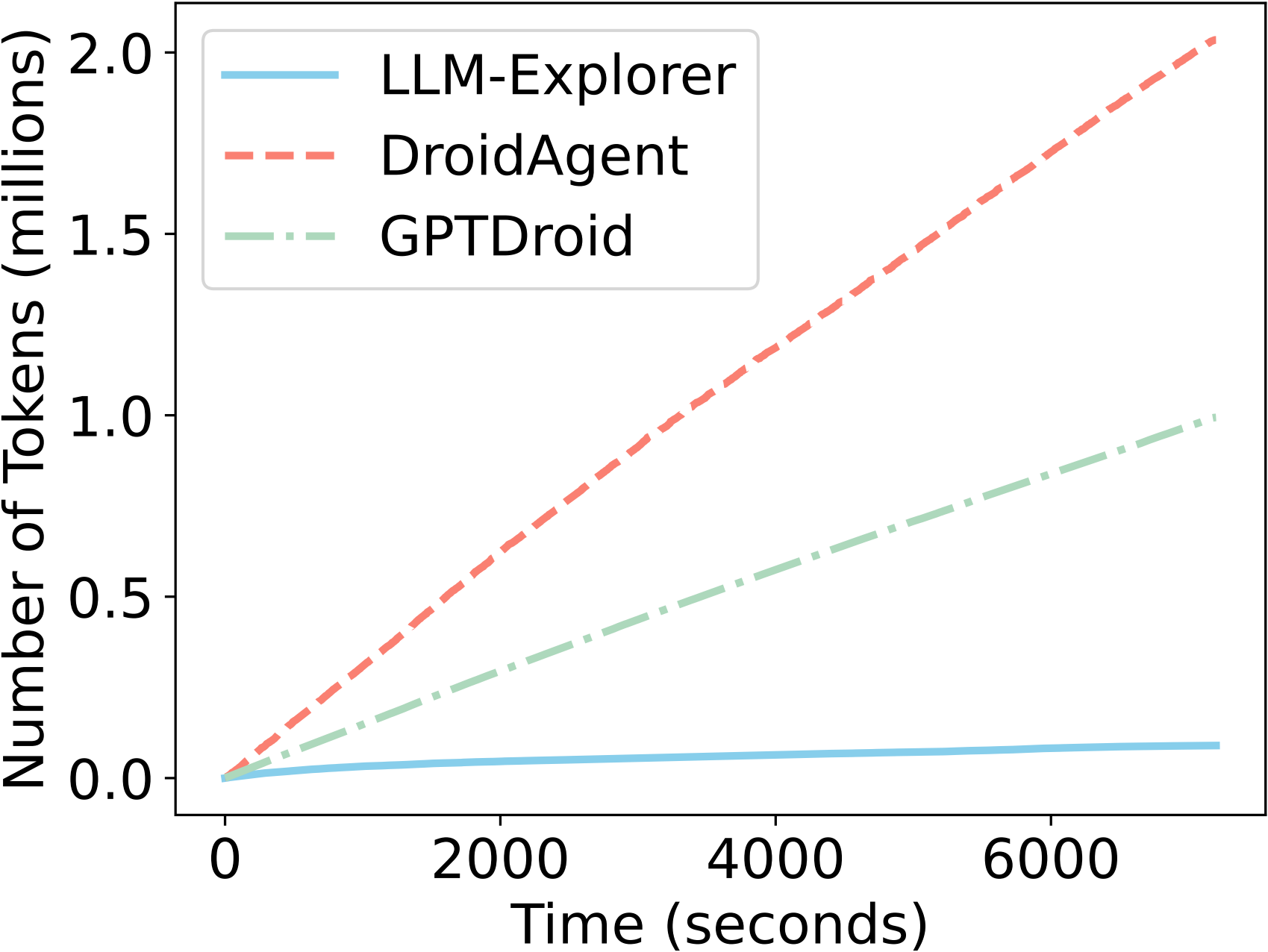}
    \caption{The progressive LLM token consumption of \name, DroidAgent, and GPTDroid.}
    \label{fig:token_by_time}
    \vspace{-0.3cm}
\end{figure}

To understand the relation between the cost of the exploration process and the complexity of the apps, we analyzed the exploration cost of \name, DroidAgent and GPTDroid on different apps. As shown in Figure~\ref{fig:cost_apps}, the cost of \name is more adaptive for different apps, with lower cost on the apps with smaller number of activities (\emph{App Launcher}, \emph{Keybord}, etc.) while higher cost on more complex apps (\emph{My Expenses}, \emph{SMS Messenger}, etc.). However, DroidAgent and GPTDroid do not clearly show this adaptivity, and won't significantly reduce costs when exploring simpler apps.
Meanwhile, the costs increased linearly with the step count in DroidAgent and GPTDroid, while the increase of \name was sublinear, as shown in Figure~\ref{fig:token_by_time}.
Such adaptivity and sublinearity come from the way \name uses LLM, which has been discussed in Section \ref{sec:detailed_usage_of_LLM}.


\subsection{Model Variation Test}
\label{eval:ablation}





\begin{figure}
    \centering
    \includegraphics[width=0.47\textwidth]{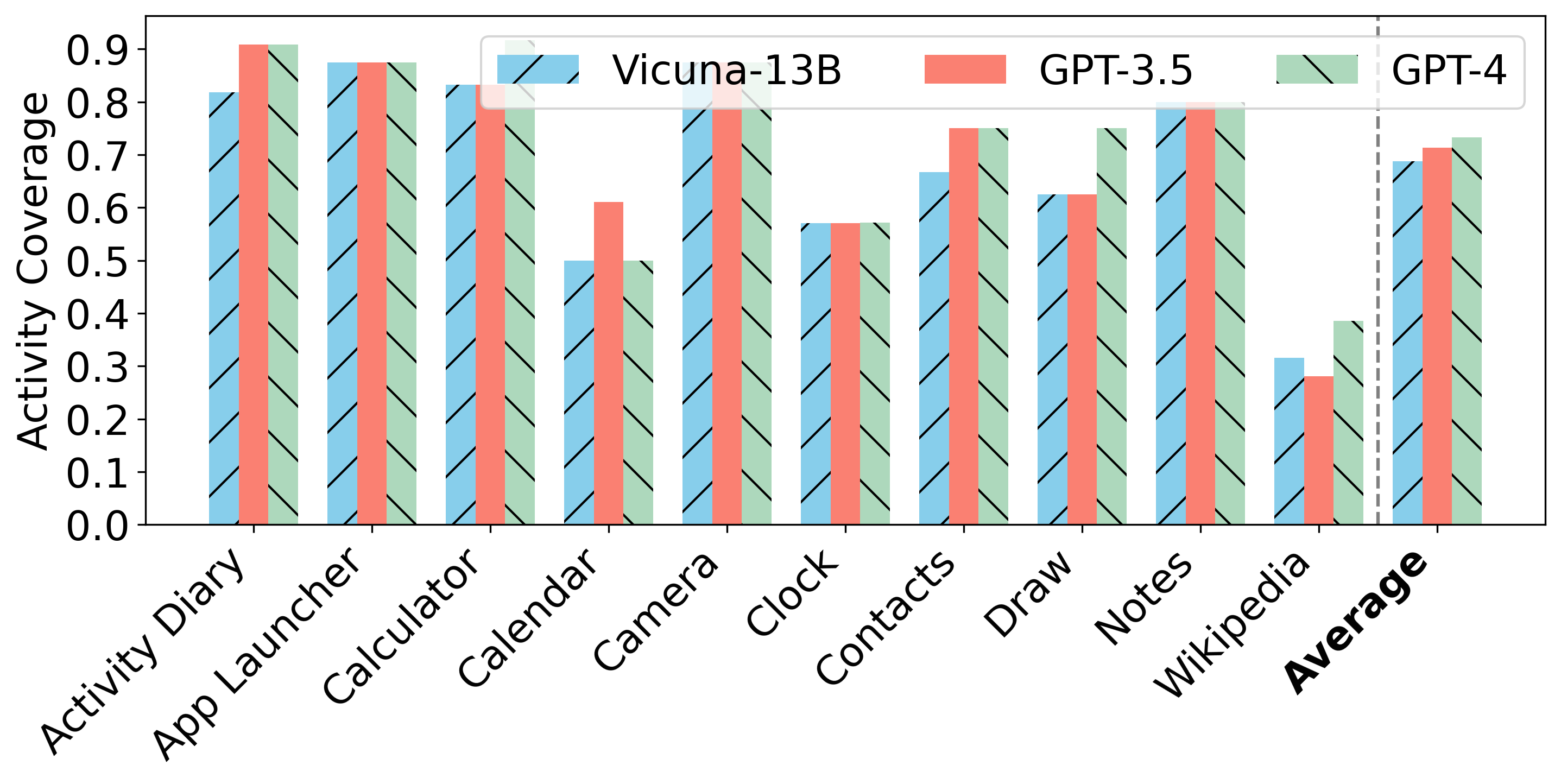}
    \vspace{-0.5cm}
    \caption{The activity coverage achieved by \name on 10 apps with different LLMs.}
    \label{fig:ablation_result}
    \vspace{-0.3cm}
\end{figure}

Given the pivotal role of the LLM (GPT-3.5) in the exploration policy of \name, we evaluate its influence by replacing it with a weaker and a stronger variant. The results are shown in Figure~\ref{fig:ablation_result}.

\textbf{Using a Smaller Local LLM.} We first tested the performance of \name with Vicuna-13B \cite{vicuna2023}. We quantized it with AWQ \cite{lin2023awq} to 4bit and deployed it on a desktop computer equipped with an NVIDIA GeForce RTX 3090 GPU.
As shown in Figure \ref{fig:ablation_result}, there were slight decreases in coverage for most of the apps when using \name with Vicuna-13B instead of the default GPT-3.5.
The reason was mainly due to the decreased quality of LLM responses.
Specifically, the smaller Vicuna-13B model may return responses with incorrect formats, invalid element ids, or wrong classifications, leading to inaccurate summarization of abstract elements.
Although we had queried the LLM for multiple times to increase the chances of good responses, the problem was not easily avoidable.
However, the Vicuna-based \name still exhibited acceptable overall performance, suggesting that \name does not heavily rely on a larger LLM and may be further improved with local model training.


\textbf{Using GPT-4.} We also evaluated the performance of \name with GPT-4.
As depicted in Figure 8, there were slight increases in coverage for most apps, attributed to the higher quality responses from GPT-4. However, despite the superior reasoning capability of GPT-4 compared to GPT-3.5, the enhancement in average coverage was not significant. Using GPT-3.5 in \name is good enough for most apps, except for some complicated apps (\eg Wikipedia) that demand more advanced knowledge management. These findings suggest that using the most powerful LLMs may not be necessary and we choose GPT-3.5 as the default choice.

\subsection{Case Studies}\label{sec:case_studies}

We further dive deeper into the succeeded and failed cases of \name to understand its advantages and limitations.

\begin{figure}
    \centering
    \includegraphics[width=0.49\textwidth]{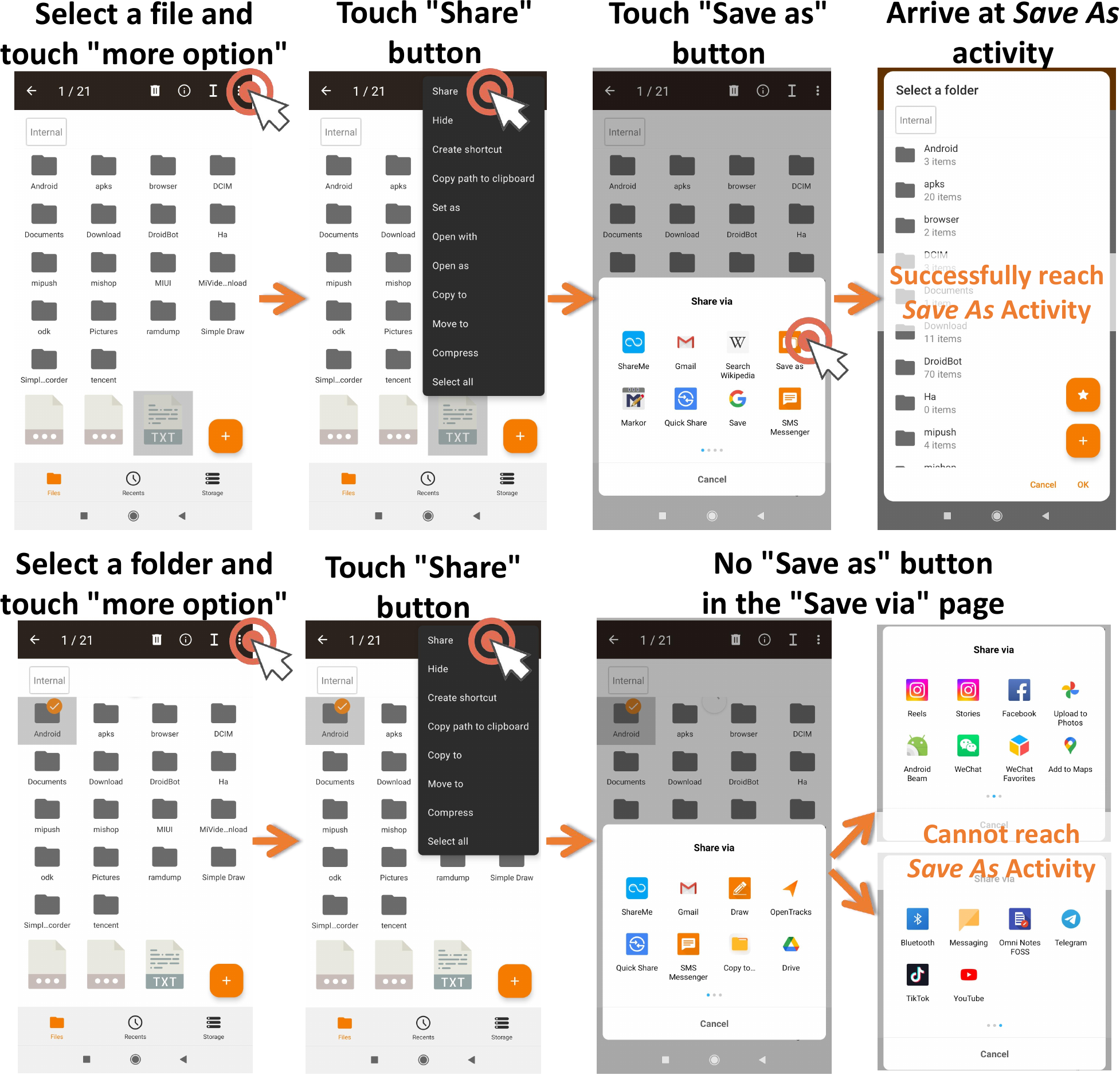}
    \caption{A case where \name outperformed the human performance. \textnormal{Top: \name successfully reached the \textit{Save As} activity by selecting a file to save. Bottom: Human users only tried to select a folder, which couldn't reach the \textit{Save As} activity.}}
    \label{fig:case_studies_success}
\end{figure}

\textbf{Cases outperforming human.}
\name achieved better coverage than human on three apps including \emph{File Manager}, \emph{Open Tracks} and \emph{SMS Messenger}. We compared the activities reached by human and \name to understand the reasons.
In Open Tracks, human participants missed the \emph{Marker Edit} and \emph{Marker Detail} activities. This was because these two activities require the creation of a marker first, but the creation of a marker fails when the GPS signal is weak.
For other activities that \name explored but human users did not, it was mainly due to human participants overlooking detailed functions, such as the \emph{Recycle Bin Conversations} activity in SMS Messenger, and the \emph{Save As} activity in File Manager. The feasible path to reach the \emph{Save As} activity is shown in the top half of Figure \ref{fig:case_studies_success}.
To reach the activity, one needs to select a file, click the ``More Options'' button, click the ``Share'' button, and finally click the ``Save as'' button on the screen. The ``Save as'' was invisible when the human user selected a folder (instead of a file) to share, and the \emph{Save As} activity was unreachable. Such a detailed function was uneasy to be noticed by human users, while automated explorers can solve this problem through extensive traversal.


\begin{figure}
    \centering
    \includegraphics[width=0.4\textwidth]{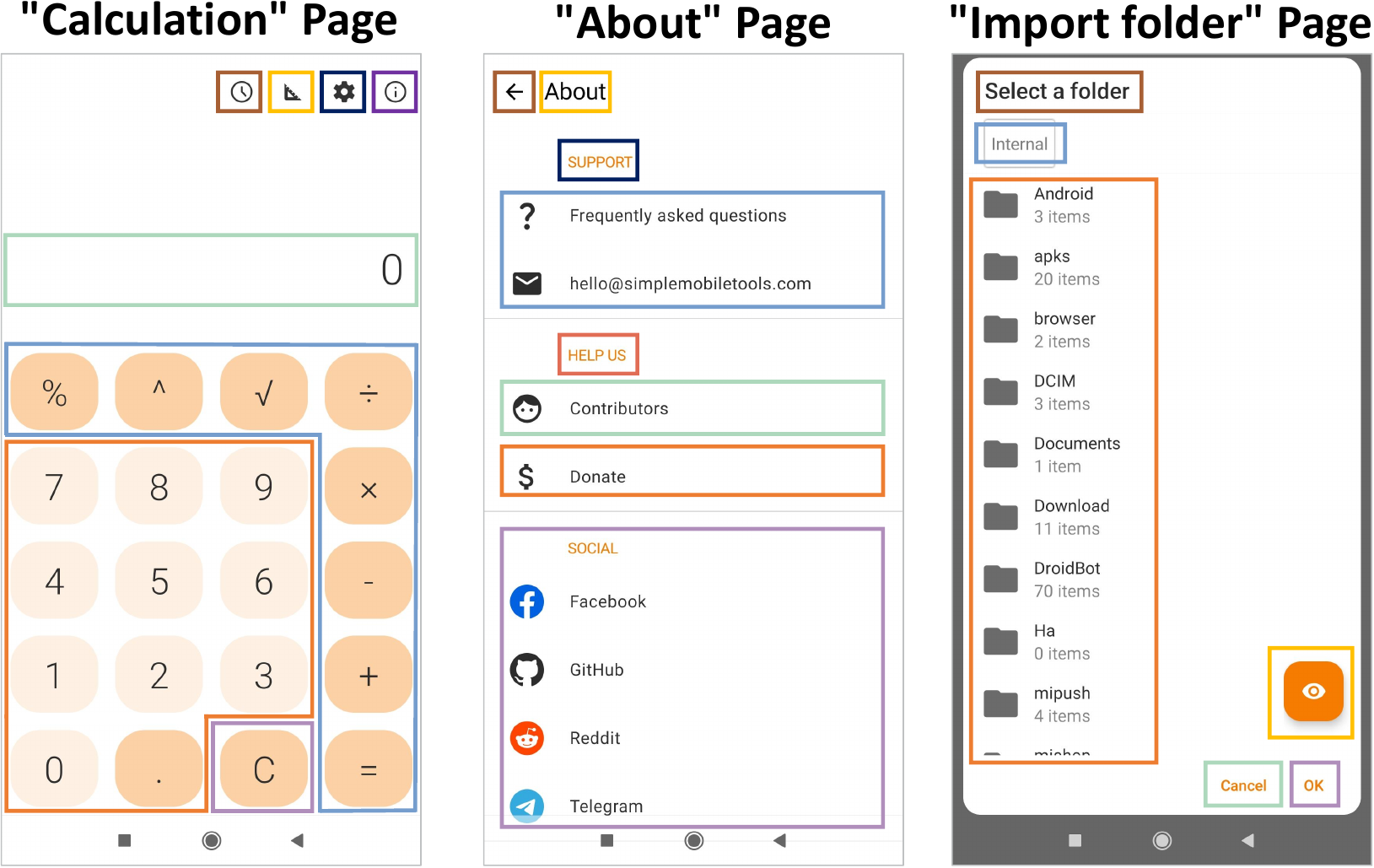}
    \vspace{-0.3cm}
    \caption{Illustrations of same-function elements determined by LLM-assisted knowledge maintenance module.}
    \label{fig:case_studies_memory}
    \vspace{-0.2cm}
\end{figure}

\textbf{Effects of LLM-assisted Knowledge Management.} The knowledge of \name could help to reduce the number of redundant actions. Figure \ref{fig:case_studies_memory} shows the same-function element groups determined by the LLM in three different GUI pages. The elements in each box were considered to have the same function and their actions were stored as one abstract action. In the \emph{Calculator} app, \name grouped the number entry buttons as a cluster, and all arithmetic symbol buttons as another cluster. In the \emph{About} page, \name was able to group several communication platforms in the \emph{SOCIAL} section as a cluster having the same function. In the \emph{Import Folder} page, \name can group several folders together. Such groupings reduced a lot of meaningless actions and action combinations.

\begin{figure}
    \centering
    \includegraphics[width=0.4\textwidth]{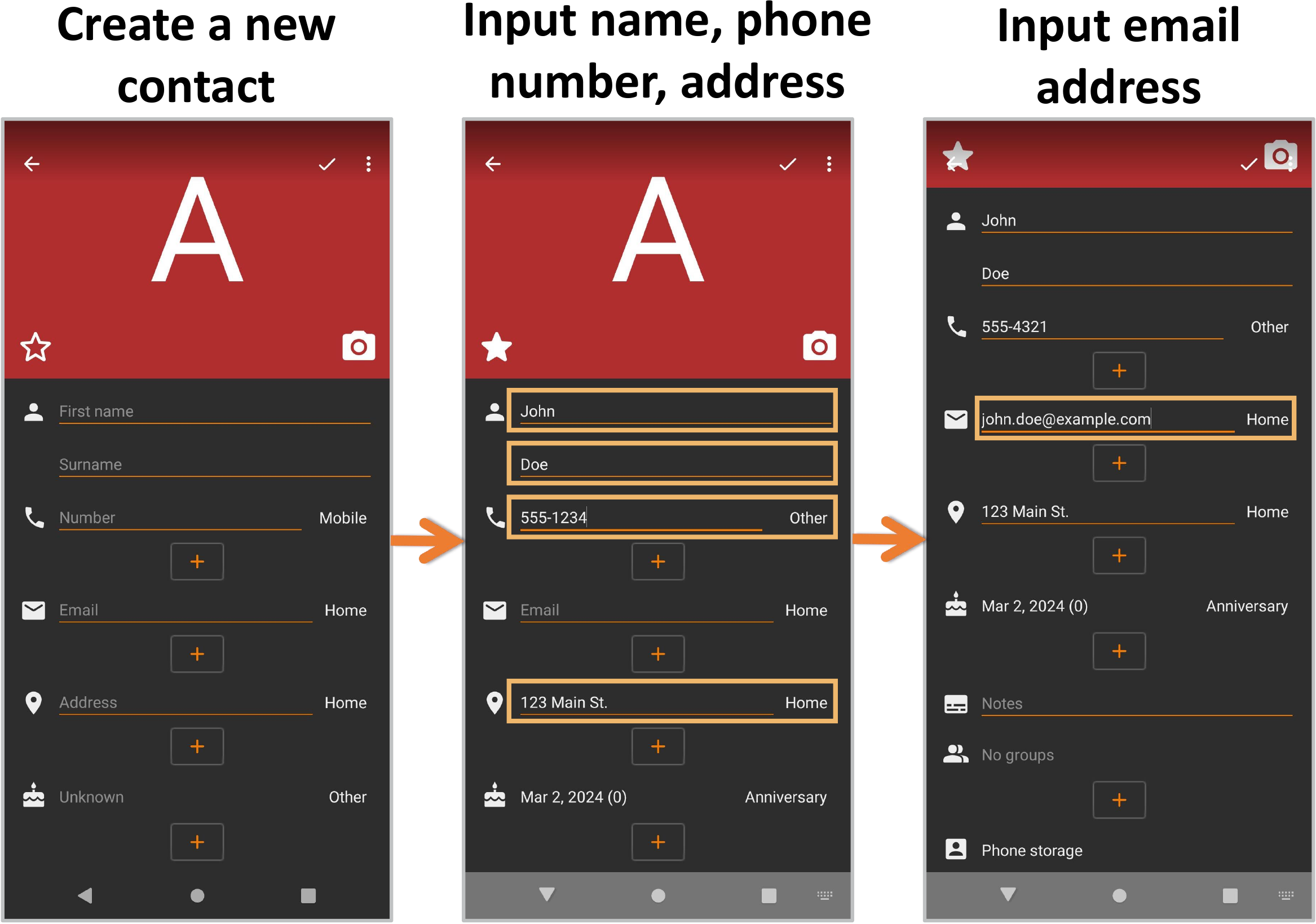}
    \vspace{-0.3cm}
    \caption{Examples of meaningful text inputs generated by \name in the \textit{Contacts} app.}
    \label{fig:case_studies_input}
    \vspace{-0.2cm}
\end{figure}

\textbf{Effects of Content-aware Input Text Generation.} \name also uses LLM to generate text input. Figure \ref{fig:case_studies_input} shows \name's input when creating a new contact in the \emph{Contacts} app. \name was able to fill in the contact's name, phone number, and address information based on the element description. After entering the user name ``John Doe'', it could generate email address information associated with the contact name. These meaningful inputs made \name easier to pass the input checks in the apps.

We also analyzed the situations where \name performed worse. The main causes are as follows.

\begin{figure}
    \centering
    \includegraphics[width=0.4\textwidth]{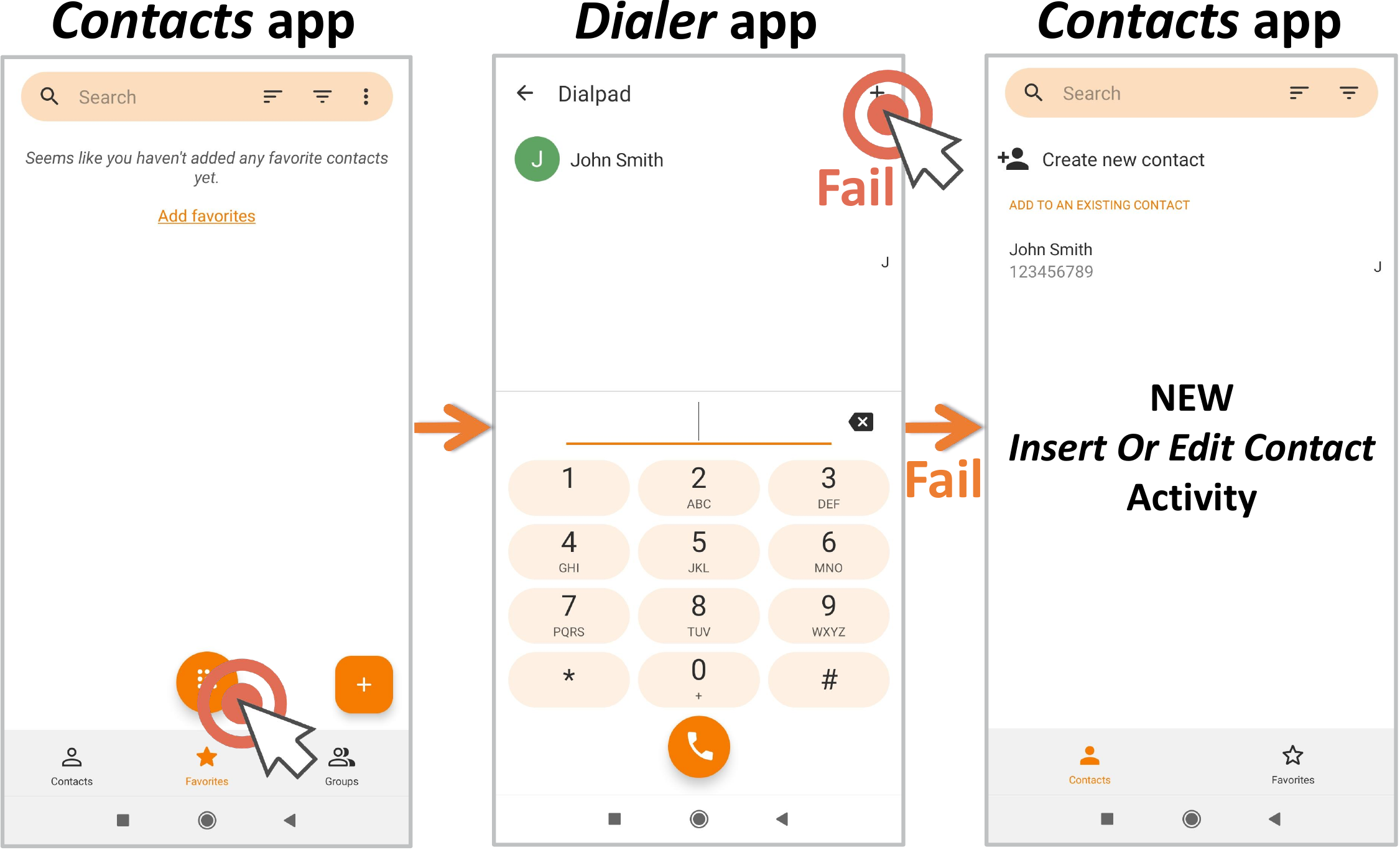}
    \vspace{-0.3cm}
    \caption{\name failed to reach the activity that required cross-app navigation.
    }
    \label{fig:case_studies_failure}
    \vspace{-0.3cm}
\end{figure}

(i) \textbf{Failed to reach activities requiring cross-app jumps.} \name performed poorly for activities that need to be navigated across apps. For example, as shown in Figure \ref{fig:case_studies_failure}, in the \textit{Contacts} app, the path to reach the \textit{Insert Or Edit Contact Activity} was to click on the \emph{Dialpad} button and then click the ``Add'' button on the \emph{Dialpad} page. However, since the \emph{Dialpad} page used by Contacts belongs to the \emph{Dialer} app (not the app under test), \name automatically returned to the \emph{Contacts} app without further exploration. This kind of failure could potentially be fixed by allowing the agent to explore outside the target app for more steps.

\begin{figure}
    \centering
    \includegraphics[width=0.4\textwidth]{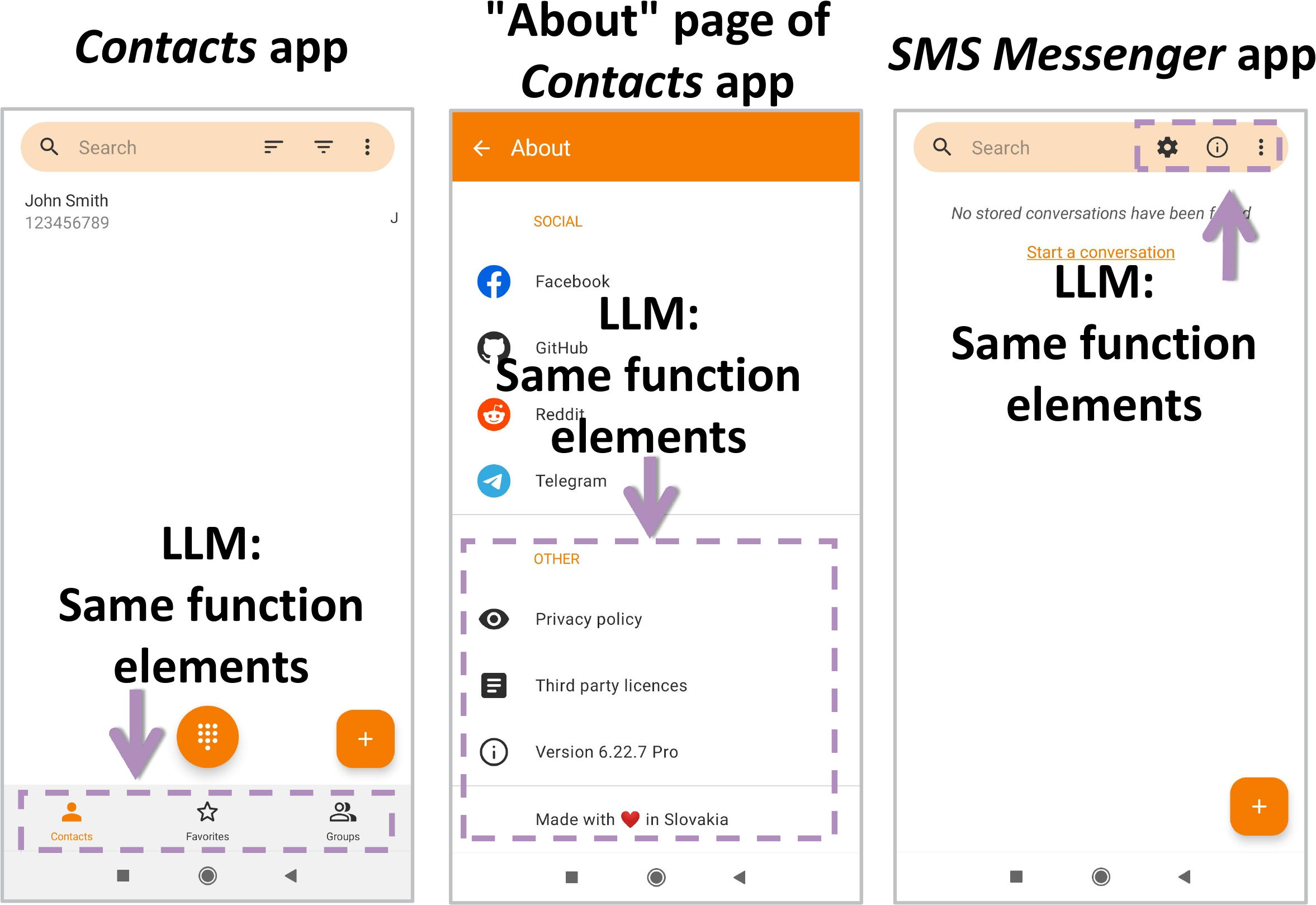}
    \vspace{-0.3cm}
    \caption{The LLM-assisted knowledge maintenance module of \name may mistakenly group actions with different functions into one abstract action.}
    \label{fig:case_studies_failure_memory}
    \vspace{-0.3cm}
\end{figure}

(ii) \textbf{Errors in maintaining the knowledge.} In the LLM-assisted knowledge maintenance module, the LLM may falsely classify elements with different functions into the same group, resulting in some useful actions being grouped into the same abstract action in the knowledge and ignored in future explorations. 
For example, when \name explored the page shown in Figure \ref{fig:case_studies_failure_memory}, the knowledge maintenance module incorrectly judged the elements in the purple box as having the same function. As a result, in the app knowledge, actions of these elements were combined as one abstract action. However, in fact, they could navigate to different GUIs. Similarly, in the \emph{About} page of the contacts app, the \texttt{touch} actions on the ``Privacy policy'' button and the ``Third-party licenses'' button were combined into the same abstract action. \name only touched the ``Privacy policy'' button and completed the exploration of this part early, missing the chance to visit the \textit{License} activity by touching the ``Third-party licenses'' button. These errors could be mitigated by using better LLMs to generate more precise abstractions.


%% file: tex/discussion.tex
\section{Discussion}\label{sec:discussion}

\textbf{Cross-app navigation.}
\name automatically navigates back when detecting a transition to another app. This design ensures that the exploration remains focused on the specified app but hinders the discovery of activities that require cross-app transitions. A potential solution is to merge the abstract interaction graphs of different apps into a unified graph, allowing \name to search for possible cross-app return paths when transitioning to another app and enabling more effective cross-app navigation and exploration.

\textbf{Utilization of VLMs in app exploration.}
Incorporating visual modalities, such as screenshots, into prompts using state-of-the-art multimodal models like GPT-4o is an effective approach to maintaining higher-quality app knowledge and has the potential to enhance system performance. Screenshots provide crucial information that may not be accurately conveyed through text alone, such as descriptions of images and icons within the interface. However, adding visual modalities typically leads to increased token consumption, which consequently raises system overhead.

%% file: tex/conclusion.tex
\section{Conclusion}

We present an LLM-based exploration agent for mobile apps.
Experiment results show that our method can achieve effective and efficient exploration, outperforming strong baselines.
By wisely choosing when and how to use LLM in the exploration process, we can significantly reduce the cost of LLMs while maintaining high exploration performance.
We believe that the synergy between LLM and well-organized domain knowledge is the key to making intelligent agents and services more affordable.

\section*{Acknowledgement}
\label{sec:acknowledgement}
This work is supported by National Natural Science Foundation of China (Grant No.62272261), Tsinghua University (AIR)–AsiaInfo Technologies (China) Inc. Joint Research Center, Wuxi Research Institute of Applied Technologies, Tsinghua University (Grant No.20242001120) and Beijing Academy of Artificial Intelligence (BAAI). Wenjie Du and Cheng Liang contributed as interns at Tsinghua University.